\documentclass{article}

\usepackage{arxiv}

\usepackage[utf8]{inputenc} % allow utf-8 input
\usepackage[T1]{fontenc}    % use 8-bit T1 fonts
\usepackage{url}            % simple URL typesetting
\usepackage{booktabs}       % professional-quality tables
\usepackage{amsfonts}       % blackboard math symbols
\usepackage{nicefrac}       % compact symbols for 1/2, etc.
\usepackage{microtype}      % microtypography
\usepackage{lipsum}		% Can be removed after putting your text content
\usepackage{graphicx}
\usepackage{natbib}
\usepackage{doi}
\usepackage{graphicx}
\usepackage{hyperref}
\usepackage{amsmath,amssymb} % define this before the line numbering.

\title{The birds need attention too: Analysing usage of Self Attention in identifying bird calls in soundscapes}

\author{Chandra Kanth Nagesh\\
	Department of Computer Science\\
	University of Colorado Boulder\\
	Boulder, CO 80309 \\
	\texttt{chandrakanth.nagesh@colorado.edu} \\
	\And
	Abhishek Purushothama \\
	Department of Computer Science\\
	University of Colorado Boulder\\
	Boulder, CO 80309\\
	\texttt{abhishek.purushothama@colorado.edu} \\
}

\date{}

\renewcommand{\undertitle}{Technical Report}
\renewcommand{\headeright}{Technical Report}
\renewcommand{\shorttitle}{The Birds Need Attention Too}

\hypersetup{
pdftitle={The birds need attention too:},
pdfsubject={audio-processing, deeplearning},
pdfauthor={Chandra Kanth Nagesh, Abhishek Purushothama},
pdfkeywords={Transformers, CNNs, Deep Learning},
}

\begin{document}
\maketitle

\begin{abstract}
	Birds are vital parts of ecosystems across the world and are an excellent measure of the quality of life on earth. Many bird species are endangered while others are already extinct. Ecological efforts in understanding and monitoring bird populations are important to conserve their habitat and species, but this mostly relies on manual methods in rough terrains. Recent advances in Machine Learning and Deep Learning have made automatic bird recognition in diverse environments possible. Birdcall recognition till now has been performed using convolutional neural networks. In this work, we try and understand how self-attention can aid in this endeavor. With that we build an pre-trained Attention-based Spectrogram Transformer baseline for BirdCLEF 2022 and compare the results against the pre-trained Convolution-based baseline. Our results show that the transformer models outperformed the convolutional model and we further validate our results by building baselines and analyzing the results for the previous year BirdCLEF 2021 challenge. Source code available at \footnote{\url{https://github.com/ck090/BirdCLEF-22}}
\end{abstract}

% keywords can be removed
\keywords{Deep Learning \and Transfer Learning \and Mel Spectrograms \and Audio Pattern Recognition \and Transformers \and Convolutional Neural Networks}

\section{Introduction}

% First paragraph
\subsection{Problem}
BirdCLEF 2022\footnote{\url{https://www.imageclef.org/BirdCLEF2022MainTask}} (Bird Cross Language Evaluation Forum) is a task in this year's edition of LifeCLEF as part of the ImageCLEF competitions. LifeCLEF aims at encouraging and innovating building computational tools for extraction and maintenance of ecological and biological knowledge to help with biodiversity conservation and accessibility. 

LifeCLEF has had many editions of BirdCLEF over the years. In this year's edition the goal is to build a system to detect and identify bird calls for Hawaiian bird species from given soundscapes. The soundscapes are expected to contain background bird songs and ambient noise.

\subsection{Motivation}
% MOTIVATION: Explain the motivation for your work; e.g., Why anyone should care? What are the desired benefits?
The organizers explain the need for such systems and hence the purpose of the task, but we summarize it below for brevity.
A large number of Hawaiian bird species(an estimated 68\%) are already extinct, and those that remain are spread across the islands, in isolated and hard to access places. Population monitoring is a method for researchers to understand how the native birds react to the changes in the environment and hence drive the conservation efforts. Given the logistical challenges that arises with population monitoring in such scenarios, scientists have turned to sound recordings to help monitor. This approach called \textit{Bio-acoustic} monitoring provides a more cost-effective, passive, low labor strategy for studying endangered bird populations. The recent advances in Machine Learning and Audio processing methods have made it possible to automatically identify bird songs for common species with ample training data. However, it remains challenging to develop such tools for rare and endangered species, such as those in Hawaii. In this problem we have to develop a model that can process continuous audio data and then acoustically recognize the bird species with this limited training data.

The previous edition BirdCLEF 2021 \footnote{\url{https://www.imageclef.org/BirdCLEF2021MainTask}} had a similar task of classifying bird sounds. The top approaches used pre-trained image based based convolution backbones such as \textit{ResNet}\cite{resnet}. Attention, specifically \textit{Self-Attention}\cite{attention-is-all-you-need} has shown great performance across various areas of applied machine learning\cite{devlin-etal-2019-bert}\cite{chen2021outperform}\cite{gong2021ast}\cite{wav2vec}. Audio Spectogram Transformer \cite{gong2021ast} is a transformer based architecture pretrained on sounds in the form of spectograms, enabling it to learn contextual representations. This provides, in the authors' opinion provides the best self-attention option that can be adapted for the specific task of bird sound classification at hand.

We propose building an AST based Attention baseline for BirdCLEF. In order to provide a comparison to a Convolution baseline for BirdCLEF, we also plan to build a model based on the pre-trained PANNs \cite{PANNs} and EfficientNet \cite{EfficientNet} architectures. This will provide us with a Attention-based baseline for the challenge, and a Convolution based baseline to compare against. This would help in deciding whether Self-Attention backbones, specifically AST would be an useful 

We also plan on evaluating the newly developed model on previous year challenge (BirdCLEF 2021) to further validate our novel proposal of an AST based pipeline for solving problems in the audio spectral domain.

\section{Related Work}

Prior edition of BirdCLEF 2021 had a similar challenge with different dataset and problem. It involved a larger set of labels (397) but had both short recordings and soundscapes available for training. 

A key factor for the challenge according to the organizers was ``Bridging the gap between \textit{high-quality training-recordings} and \textit{ soundscapes with high-ambient noise levels} was one of the most challenging tasks in audio event recognition". 

Common methods of the top submissions, were use of \textit{mel scale spectograms} as inputs and use of \textit{mix-up} for data-augmentation. For post-processing with \textit{bagging} and \textit{thresholding}, \textit{location filtering}, \textit{decision tree aggregation} of scores and metadata were also impactful \cite{KahlEtAl:CLEF-2021}. There are 7 system descriptions that are available in the working notes \cite{CLEF-2021} and are discussed under their specific tags. \\ 

\noindent\textbf{Convolution neural network based architectures} such as  \cite{shawn2016}, \cite{PANNs}, \cite{efficient-net}, and \cite{Ford2019ADR} were studied. Although, these systems have shown state-of-the-art performance on audio tasks, they have typically performed well when there were large pool of audio data with a diverse labels. Our work differs in that we will be training this model for a comparative study on a specific task which involves lesser quantity and diversity of data. \\
% Nickname the systems since they will be rementioned. 

\noindent\textbf{Audio Transformer architectures} such as \cite{gong2021ast} and \cite{li2022audiotagging} were studied. Attention based models and especially Transformer architectures such as AST, provide extensibility and adaptability. Through our literature review we have not seen any application of these architectures for previous challenges. Hence our work differs in the way that we hope to apply the same to the BirdCLEF 2022 challenge.

\noindent\\
\textbf{Combating Weak-supervision} \cite{HenkelEtAl:CLEF-2021} (Second place solution in last year challenge), \cite{CondeEtAl:CLEF-2021}, \cite{ShugaevEtAl:CLEF-2021} and \cite{MurakamiEtAl:CLEF-2021} were studied. The above systems provide complex and cumbersome methods for combating weak supervision which is present in the dataset and the task. Due to the nature of the challenge, there are no clear answer to effectiveness of varied convolutional architectures or the compounding effects of different processing techniques. \\

In addition, all these methods utilize convolution backbones. Our proposal utilizes transformer architecture to build a general model to which additional processing techniques can be supplemented to combat weak supervision.

\section{Methods}
% Methods - describe the implementation of your proposed idea (e.g., features, algorithm(s), training overview) so that:
% A reader could reproduce your set-up
% A reader understand why you made your design decisions
% todo: make a table of Mel hyperparams - Done
% todo: table of model hyperparams - not needed as we have mentioned most of them through out this section 
% todo: cite pytorch lightning - Done
\subsection{Data}
We train the models on data from BirdCLEF 2022 and validate them, we also further justify our findings by training and validating on the data from BirdCLEF 2021.

The BirdCLEF 2022 challenge consists of bird calls captured by citizen scientists for the Xeno-canto network. It training data consists of 15,000 bird calls convering almost 152 species from across the worlds. The BirdCLEF 2021 challenge on the other hand consists of over 62,000 bird calls from all around the world, covering 397 bird classes. The classes of birds that were included for example were Akiapo, Aniani and Barbet to name a few.\\
% Add appendix with class distribution - will be relevant for Macro vs Samples discussion.
Since there is insufficient validation and test data provided at the current time, we build our models for the task of sound classification for the 152.

The audio data are provided in the corresponding folders representing the bird names. Each of the audio recordings are varied in length, some are longer than 10 seconds while others are just 2 seconds. The same applies for both the challenges as well. From here, we pick a maximum of 40 audios from each of the bird classes and treat that as our data set. Therefore, we have 4898 audio files for the BirdCLEF22 challenge and 15614 audio files for the BirdCLEF21 challenge. Then we split the data into training and validation splits using the Stratified Shuffle Split.

\begin{figure}[h!]
\centering
  \includegraphics[width=0.5em]{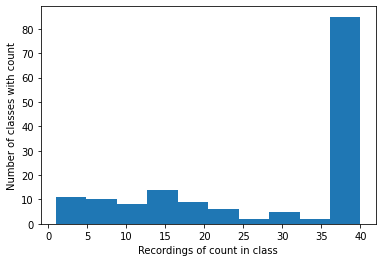}
  \caption{Distribution of counts of recordings for classes}
  \label{fig:dist}
\end{figure}

We select upto 40 audios for each class, but many of the classes have very few recordings to begin with. A histogram for the distribution of the count of recordings per class \ref{fig:dist}.

These audio files paths are part of a data frame which is created for each of the bird classes in both the challenges. The audio files are read using the librosa framework provided by PyTorch \cite{PANNs}.

\subsection{Features/Data set processing}
From the recorded audio data set, given as part of the challenge we will be considering all the audio signals and chopping them into 10 second intervals with overlap and converting them into \textit{Mel Spectrogram}. The \textit{mel scale} is a lograthimic scale where equal distances on the scale have the same perceptual distance. $1000 Hz = 1000 Mel$. As frequency increases, the interval, in hertz, between mel scale values increases. In order to extract Mel spectrograms we first extract the Short Time Fourier Transforms, this gives us amplitudes, which we then convert to decibels and then convert the frequencies into the Mel scale. 

\begin{figure}[h!]
\centering
  \includegraphics[width=0.6\linewidth]{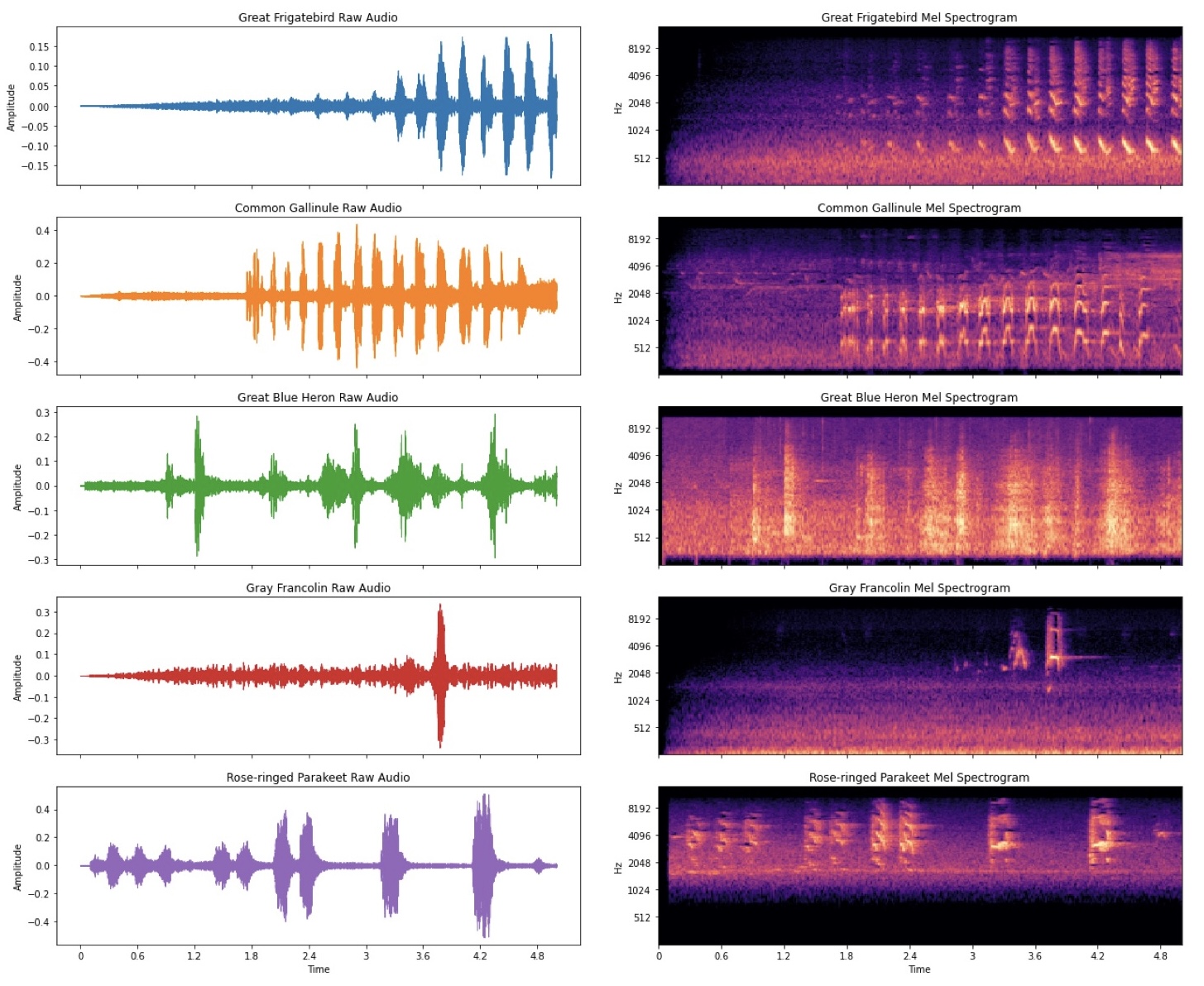}
  \caption{Simple wave of the Bird calls and their corresponding Mel Spectrogram}
  \label{fig:chirp}
\end{figure}

Figure \ref{fig:chirp} shows the amplitude of the bird sound as a function of time. Now, we can convert the audio wave into mel spectrogram and this will yield us the right part of \ref{fig:chirp} i.e. their corresponding Mel spectrograms for 5 of the bird represented in our population as part of the challenge.
We set the following parameters to create the Mel spectrograms as shown in \ref{fig:chirp} from the given waveform data. We tested multiple alternatives and came up with these final parameters. The Mel spectrograms were generated with the following parameters.

\begin{table}[h!]
    \centering
    \begin{tabular}{|c|c|}
        \hline
        Mel Filterbanks & 128 \\
        \hline
        Minimum frequency & 20Hz \\
        \hline
        Maximum frequency & 16000HZ \\
        \hline
        Sampling rate & 32000 \\
        \hline
        Fast Fourier Transforms (FFT) size & 512 \\
        \hline
    \end{tabular}
    \caption{Mel Spectrogram parameters}
    \label{tab:melparams}
\end{table}

The images which are as a result the output of the Mel Spectrogram conversion are set to be 224x224. We tried with the images sizes of 512x512 but easily ran out of memory, as the models themselves are large.

\subsection{Model/Algorithms}
\textbf{Pre-trained Convolution}\\
The Pretrained Audio Neural Networks (PANN) \cite{PANNs} are a set of prominent image classification models like ResNet, VGG, MobileNets that have been trained on the YouTube-100M dataset. This model has been trained for the audio tagging task, which is in essence a similar task as our bird call identification. We plan on fine tuning the ResNet38 model (which has shown state-of-the-art performance on the dataset) for the bird call identification challenge. This modified Resnet architecture consists of 16 basic hidden blocks of varying convolutions and 4 maximum pooling blocks, followed by 2 fully connected layers with the final layer consisting of a sigmoid activation function. The ResNet38 model has shown to perform on-par with their novel architecture the Wavegram-Logmel-CNN.\\
\\
\textbf{Pre-trained Transformer}\\
The proposed Audio Spectrogram Transformer model will be trained in a similar fashion. However, the important consideration here is that the Transformer model will require the audio spectrograms to be divided into multiple chunks in sequence which is similar to the following paper \cite{image1616}. So, we divide the 2D audio spectrogram into 16x16 patches with overlap and project then into a linear projection as a sequence of 1D patch embedding of the size 768. Each of this embedding is then added a positional token similar to any other transformer based model, which means we are adding learnable positional embedding. The transformer encoder used in this model has 12 layers, and 12 heads. Finally an additional classification token in appended to the sequence. The output of these embedding layers is then fed as the input to a standard transformer architecture. We plan on fine-tuning the hyperparameters for this step. The output of the transformer model will then be passed through a linear layer with sigmoid activation for the classification output.\\
\\
\textbf{Pre-trained Convolution Baseline}\\
We also compare the model performances against a pre-trained convolutional baseline. This is the State-of-the-Art image pre-trained model, EfficientNet-B0\cite{EfficientNet} from the MBConvNet family. This architecture is our baseline for models using convolution operation. The model consists of 9 convolutional blocks that are of varying sizes that each transform the given data. Starting from getting the input images of the size 224, architecture comprises of 7 MBConv blocks of strides 3 and 5. Finally, a convolution block with a pooling and linear layer. We then set the last layer of this pre-trained model with sigmoid activation to reflect the number of classes that are present in our datasets. \\

The complete description of the modelling framework will be presented in the Experimental Design section.

\subsection{Post processing}
Since the goal of the work is to decide between Convolutional and Self-Attention backbones, we do not perform any sort of post processing. \cite{CondeEtAl:CLEF-2021} \cite{HenkelEtAl:CLEF-2021} and \cite{MurakamiEtAl:CLEF-2021} suggest various post processing strategies and claim to improve the leader board scores. Especially, \cite{MurakamiEtAl:CLEF-2021} uses gradient boosted decision trees to eliminate false detection. \cite{CondeEtAl:CLEF-2021} employed latitude and longitude information for detecting if birds ever co-occur in the same region (since we have a soundscape) and call or no-call detection giving a performance improvements of 0.008-0.01 on the public leader board. These methodologies can always be used in addition to the AST backbone designed and used in our experiments.

\subsection{Training Overview}
We are currently planning to rapidly prototype and train on Nvidia Tesla P100 GPUs offered by Kaggle. We fine-tune the hyper parameters i.e. learning rate, epochs for training. Our preferred choice of optimization algorithm for the training is the popular Adam algorithm. The number of classes were set to 152 (BirdCLEF22) and 397 (BirdCLEF21).

The final set of hyper-parameters that were used in this project are as follows:

\begin{table}[h!]
    \centering
    \begin{tabular}{|c|c|c|}
        \hline
        Learning Rate & 1e-4 \\
        \hline
        Total Epochs & 40 \\
        \hline
        Training Batch Size & 10 \\
        \hline
        Validation Batch Size & 2 \\
        \hline
        Regularization & Early Stopping (patience = 10 epochs) \\
        \hline
        Optimizer & Adam \\
        \hline
        Learning Rate Scheduler & Cosine Annealing \\
        \hline
    \end{tabular}
    \caption{Training Hyperparameters for both AST and EfficientNet models}
    \label{tab:hyperparams}
\end{table}

We log all the results from the training to Weights and Biases\cite{wandb}, from where we report the results in the Experimental Results section. We use the popular PyTorch Lightning \cite{falcon2019pytorch} framework for the entire project.
% Experimental Design - describe 1-2 experiments or analyses you plan to conduct in order to demonstrate/validate the target contribution(s) of your work. Your description should be detailed enough so that a reader could reproduce it. Your description should include the following for each experiment:
% Main purpose: 1-3 sentence high level explanation
% Evaluation Metric(s): which ones will you use and why?

% Experimental Design (improve upon the material from your project outline)

\section{Experimental Design}

Our experimental design involves three neural network models plus common pre-processing steps. We will be performing pre-processing on the given data and converting the audio signals into images, each 10 seconds long.

Our design involves training six models, with \textbf{three} varied \textit{architectures} and \textbf{two} different data sources, they can be seen in Table \ref{tab:models}. 

\begin{table}[h!]
    \centering
    \begin{tabular}{|c|c|c|}
    \hline
    \textbf{Model Name} & \textbf{Architecture} & \textbf{Dataset}\\
        \hline
         AST-Bird-22& AST & BirdCLEF 2022 \\
         \hline
         PANN-Bird-22& PANN & BirdCLEF 2022 \\
         \hline
         EfficientNet-Bird-22& EfficientNet & BirdCLEF 2022 \\
         \hline
         AST-Bird-21& AST & BirdCLEF 2021 \\
         \hline
         PANN-Bird-21& PANN & BirdCLEF 2021 \\
         \hline
         EfficientNet-Bird-21& EfficientNet & BirdCLEF 2022 \\
         \hline
    \end{tabular}
    \caption{Architecture and Datasets for Models in our experimental design}
    \label{tab:models}
\end{table}

\subsection{Main Purpose}
The purpose of our design is to analyze performance of the AST backbone (pretrained self-attention) for bird sound classification task, and it's ability to effectively replace convolutional backbone. For the analysis to be complete our experiments include comparison with pretrained convolutional backbone baseline. One convolutional baseline should have been sufficient comparison, but the difficulty in adapting the PANN baseline (details discussed in results) encouraged us to add the alternative convolutional baseline of EfficientNet\cite{efficient-net} to our experiments. \\

\subsection{Model Design}
\begin{figure}[h!]
\centering
  \includegraphics[width=0.9\linewidth]{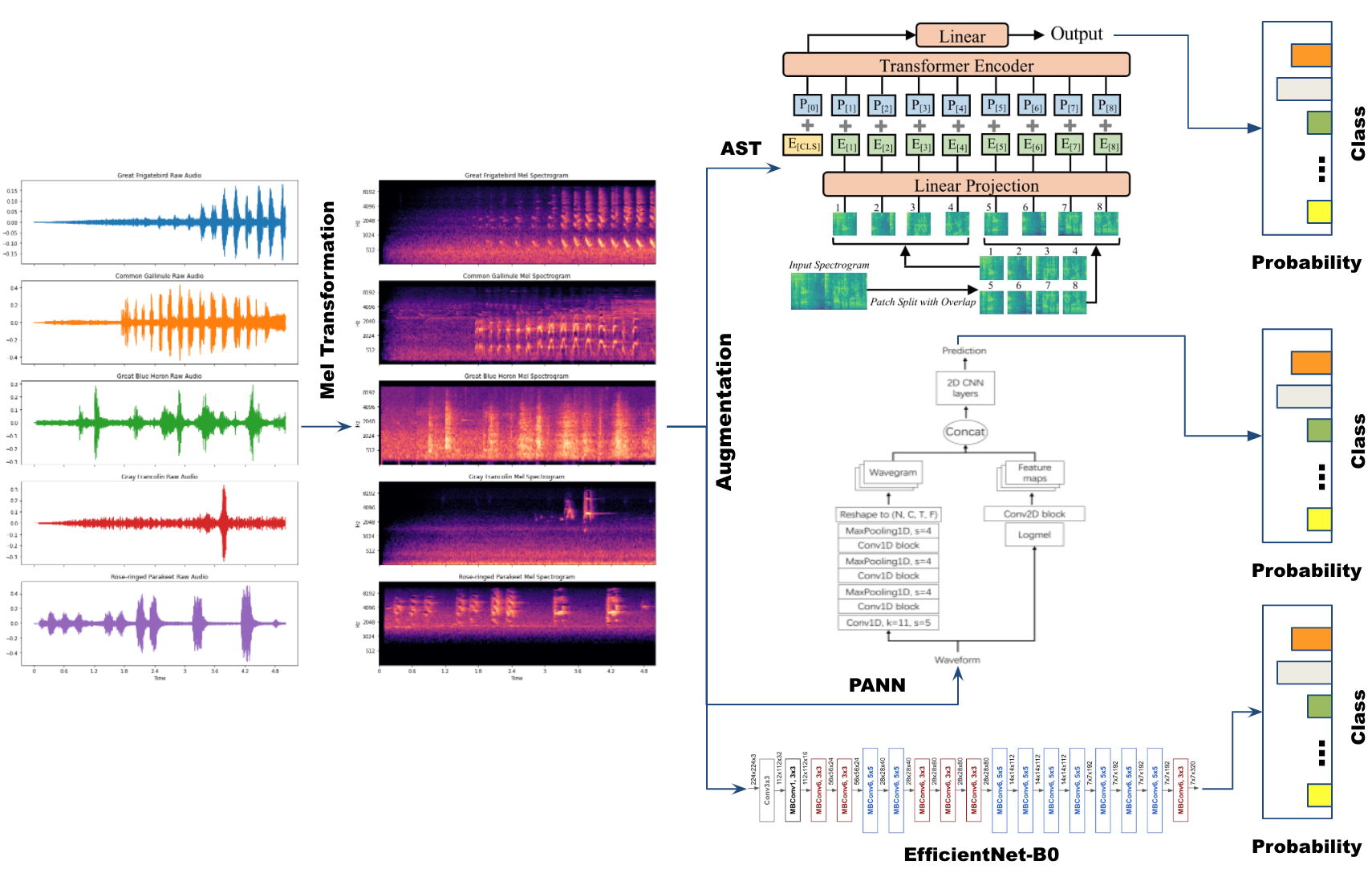}
  \caption{Our model pipeline}
  \label{fig:pipeline}
\end{figure}
Figure \ref{fig:pipeline} represents the pipeline that has been designed for this project. As mentioned in the previous section, we take the raw audio files convert them to Mel spectrograms, perform simple audio augmentation like re-sampling multiple seconds in the intervals if the audio signal is shorter. Then we feed this pipeline into the AST, PANN and EfficientNet-B0 model. The output of these models are class predictions that we show-case in this experiment.

\subsection{Evaluation Metric}
The organizers use a metric \textit{close} to the \textit{Macro F1 Score}. They define it as follows:\\
% \say{\textit{Given the amount of audio data used in this competition it wasn't feasible to label every single species found in every soundscape. Instead only a subset of species are actually scored for any given audio file.
% After dropping all of the un-scored rows we technically run a weighted classification accuracy with the weights set such that all of the species are assigned the same total weight and the true negatives and true positives for each species have the same weight. For offline cross validation purposes, the macro F1 is the closest analogue to the actual metric.}}
The F1 score is defined as:
\begin{align*}
    F1 \text{score} = 2 * \frac{\text{precision * recall}}{\text{precision + recall}}
\end{align*}
Where, $\text{Precision} = \frac{\text{TP}}{\text{TP + FP}}$ and $\text{Recall} = \frac{\text{TP}}{\text{TP + FN}}$. $TP$ is the total true positives, $FP$ is the false positives and $FN$ is the false negatives.

When you have multiple classes, the question how the F1 is calculated. With \textit{Macro F1} the F1 score is calculated per label and an \textit{unweighted mean} is calculated. We also report the \textit{Samples} F1 score.

% Experimental Results: this should include the raw results, what general trends are observed, and insights/speculations into why your results may be turning out the way they are. Also include at least one paragraph explaining what questions are not fully answered by your experiments and natural next steps for this direction of research.
%todo: we have attempted PANN training mention the downfall and explain
\section{Experimental Results}
We report the results and perform detailed analysis from our 6 experiments on the challenges as follows.

\subsection{PANN experiment}
Out of our six planned and we trained only 5 models except PANN-Bird-21, one of them unsuccessfully, namely PANN-Bird-22.

For our PANN-Bird models we re-used the original code from the PANNs paper available publicly\footnote{\url{https://github.com/qiuqiangkong/audioset_tagging_cnn}}. Our attempts at training PANN-Bird-22 failed as our model did not converge even with attempts to tune with some of the hyperparameters. However, due to this reason we do not attempt to train PANN-Bird-21, and provide a \textit{strong alternative to convolution pre-trained models} which is the EfficientNet pre-trained backbone to provide contrast to the AST Backbone. Hence, making sure that we provide the right comparison and contrast to our experiments.

\subsection{Analysis}
We report the scores for the models (Except \textit{PANN-Bird-22}) in both ``Macro" and ``Samples". We will discuss the implication of the differences in the values between them later.

\begin{table}[h!]
    \centering
    \begin{tabular}{|c|c|c|c|}
        \hline
        Model & Training Macro F1 & Validation Macro F1 & Total Time \\
        \hline
        EfficientNet-Bird-22 & 0.0311 & 0.0028 & 12897.5 \\
        \hline
        PANN-Bird-22 & 0.005 & 0.003 & 6600 \\
        \hline
        AST-Bird-22 & \textbf{0.06266} & \textbf{0.0049} & 9084.1\\
        \hline
    \end{tabular}
    \caption{`Macro' average F1 results from the Experiments for BirdCLEF 22}
    \label{tab:macrof122}
\end{table}

\begin{table}[h!]
    \centering
    \begin{tabular}{|c|c|c|c|}
        \hline
        Model & Training Samples F1 & Validation Samples F1 & Total Time \\
        \hline
        EfficientNet-Bird-22 & 0.478 & 0.3249 & 12897.5 \\
        \hline
        AST-Bird-22 & \textbf{0.9158} & \textbf{0.5046} & 9084.1 \\
        \hline
    \end{tabular}
    \caption{`Samples' average F1 results from the Experiments for BirdCLEF 22}
    \label{tab:samplesf122}
\end{table}

\begin{table}[h!]
    \centering
    \begin{tabular}{|c|c|c|c|}
        \hline
        Model & Training Loss & Validation Loss & Total Time \\
        \hline
        EfficientNet-Bird-22 & 0.010 & 0.022 & 12897.5 \\
        \hline
        AST-Bird-22 & \textbf{0.0004} & \textbf{0.0155} & 9084.1 \\
        \hline
    \end{tabular}
    \caption{Training and Validation losses from the Experiments for BirdCLEF 22 models}
    \label{tab:losses22}
\end{table}

From the above table \ref{tab:macrof122} and \ref{tab:samplesf122} we can clearly see that the AST (Audio Spectrogram Transformer) model has \textbf{outperformed} both the convolution based model. The AST model uses Vision Transformers \cite{ViT} which is the transformer architecture that uses attention mechanisms to attain state-of-the-art performance on image recognition tasks (ImageNet, CIFAR-100 etc.). Since, we are using the same model, which is now pre-trained on music spectrograms we see quite a big improvement in the overall F1 score of the model. \textit{This goes to show that using pre-trained models will especially be useful in such scenarios}. Finally, we can now compare the transformer based model against the convolutional model and see that the model has performed far lower in terms of F1 score than the model using transformers. This can be \textit{honed to the fact that the attention mechanism used in the pre-trained AST model is pertinent in identifying the subtle bird calls and their respective frequency curves}. Since, the AST model can be seen in generality to split the data into 16x16 size images and them send them to the transformer architecture, we can see that all the small details of the model are captured and this improves the overall accuracy of the model.

\begin{figure}[h!]
\centering
  \includegraphics[width=0.8\linewidth]{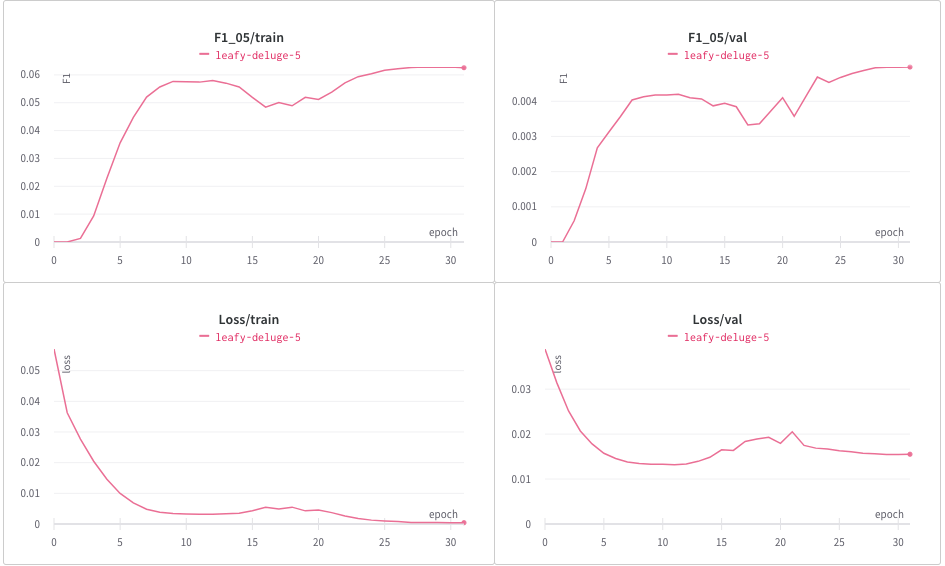}
  \caption{AST-Bird-22 F1 scores and Loss curves}
  \label{fig:AST22}
\end{figure}

\begin{figure}[h!]
\centering
  \includegraphics[width=0.8\linewidth]{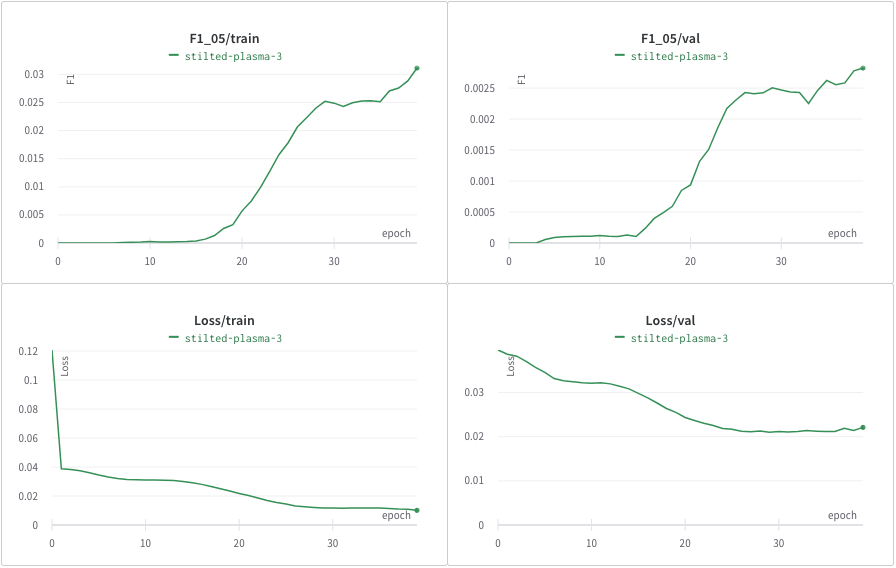}
  \caption{EfficientNet-Bird-22 F1 scores and Loss curves}
  \label{fig:EB22}
\end{figure}

From the above graph \ref{fig:AST22}, we can see that the model stopped training at 31 epochs. This is the early-stopping regularization that we used in the pipeline, that makes sure that the network stops training when parameter updates no longer begin to yield improvements on the validation set. Since, it is working by restricting the optimization procedure to smaller parameter space. We can also see that the model starts to converge after only 10 epochs as opposed to the other models that we showed. The loss curves \ref{tab:losses22} on the other hand keep steadily decreasing until epoch 8 for the training set and then decreases in small increments until the model no longer tends to learn more.

Similarly from graph \ref{fig:EB22} we can see the training curves for the Convolutional baseline model. We can clearly see how the training F1 curve starts to only converge from epoch 21 on and reaches a far lower F1 score than the transformer based model. The validation F1 also behaves the same way. The loss on the other hand  \ref{tab:losses22} first steadily decreases and reaches a plateau at around 35 epoch, while the loss curve from the training steps keep decreasing. This model trained for all 40 epochs whereas the AST model trained for 31 epochs.

We can also talk a bit about the training times taken by both the models, we can clearly see that the AST model, with the same set of hyper-parameters took lesser time to train than the EfficientNet-B0 model. This can be honed to the fact \textit{that the model converged a lot quicker than the convolution baseline model, hence kicking in the early stopping regularization}

From the above analysis we can clearly state that the transformer based model clearly outperformed the convolutional model by a wide margin. Now we can analyze the same models on the previous years challenge to prove the robustness of the AST model on such audio recognition challenge. We expect to see a similar behaviour in the performance when the models are run against the BirdCLEF 2021 challenge.

We now report the results and perform detailed analysis from our experiment on the BirdCLEF 2021 challenge.\\

\begin{table}[h!]
    \centering
    \begin{tabular}{|c|c|c|c|}
        \hline
        Model & Training Macro F1 & Validation Macro F1 & Total Time \\
        \hline
        EfficientNet-Bird-21 & 0.0062 & 0.0003 & 22292.5 \\
        \hline
        AST-Bird-21 & \textbf{0.02453} & \textbf{0.0007} & 19190.0 \\
        \hline
    \end{tabular}
    \caption{`Macro' average F1 results from the Experiments for BirdCLEF 21 models}
    \label{tab:macrof121}
\end{table}
% \vspace{-3.5em}
\begin{table}[h!]
    \centering
    \begin{tabular}{|c|c|c|c|}
        \hline
        Model & Training Samples F1 & Validation Samples F1 & Total Time \\
        \hline
        EfficientNet-Bird-21& 0.478 & \textbf{0.380} & 22292.5 \\
        \hline
        AST-Bird-21 & \textbf{0.9833} & 0.2068 & 19190.0 \\
        \hline
    \end{tabular}
    \caption{`Samples' average F1 results from the Experiments for BirdCLEF 21 models}
    \label{tab:samplesf121}
\end{table}
% \vspace{-3.5em}
\begin{table}[h!]
    \centering
    \begin{tabular}{|c|c|c|c|}
        \hline
        Model & Training Loss & Validation Loss & Total Time \\
        \hline
        EfficientNet-Bird-21 & 0.005 & 0.0129 & 22292.5 \\
        \hline
        AST-Bird-21 & \textbf{0.0009} & \textbf{0.0126} & 19190.0 \\
        \hline
    \end{tabular}
    \caption{Training and Validation losses from the Experiments for BirdCLEF 21 models}
    \label{tab:losses21}
\end{table}
% \vspace{-2em}

\begin{figure}[h!]
\centering
  \includegraphics[width=0.8\linewidth]{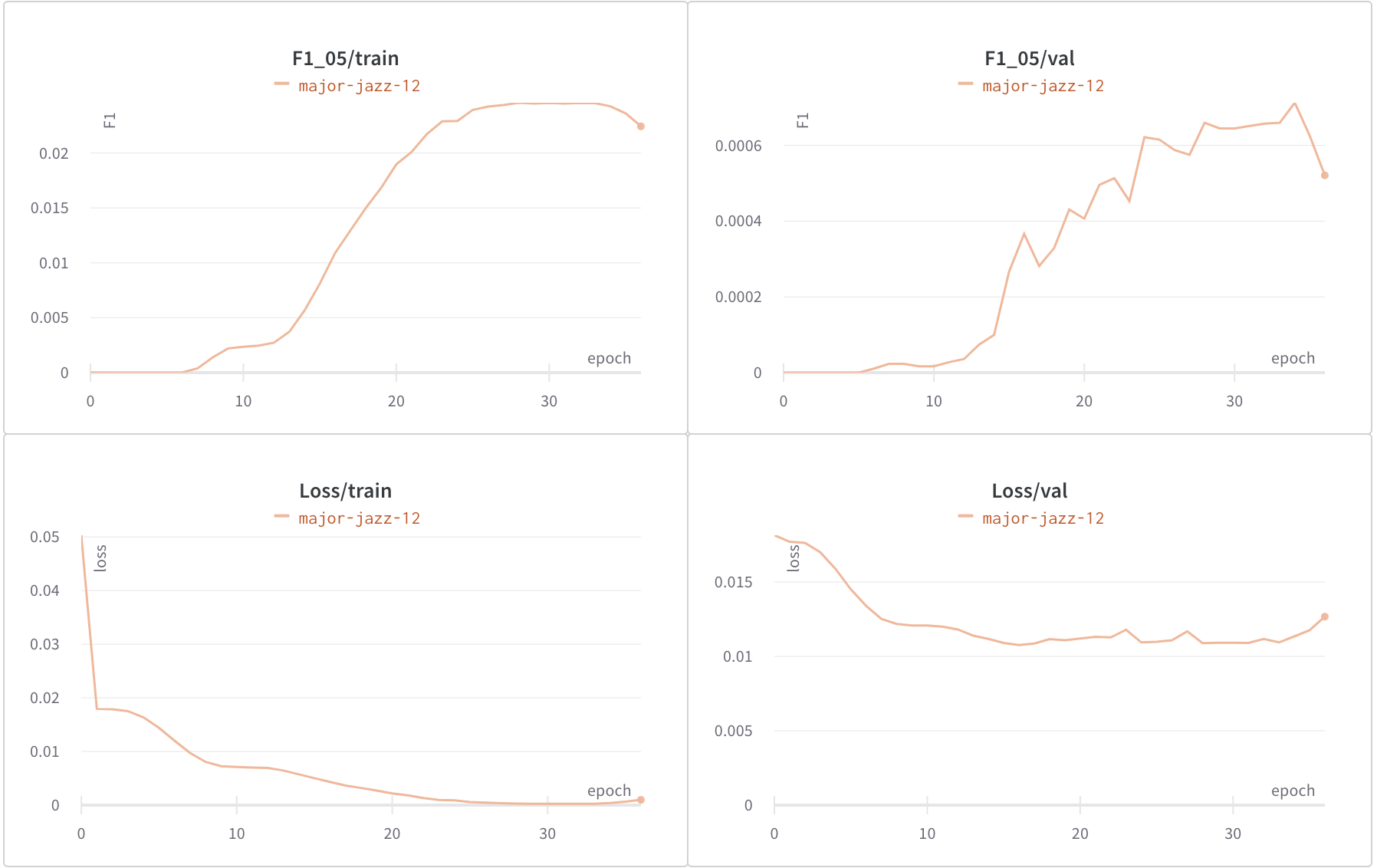}
  \caption{AST-Bird-21 F1 and Loss curves}
  \label{fig:AST21}
\end{figure}

\begin{figure}[h!]
\centering
  \includegraphics[width=0.8\linewidth]{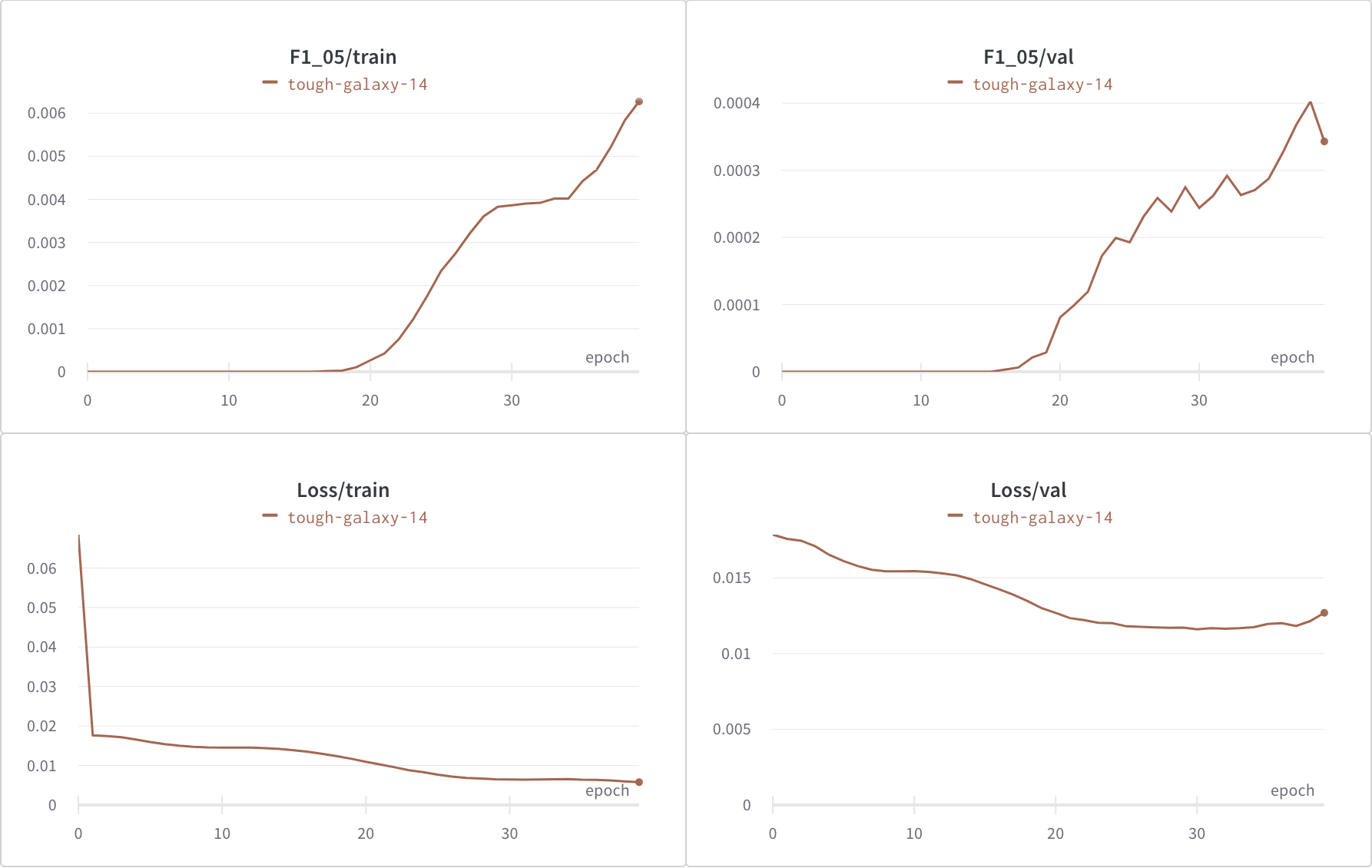}
  \caption{EfficientNet-Bird-21 F1 and Loss curves}
  \label{fig:EB21}
\end{figure}

From \ref{tab:macrof121} and \ref{tab:samplesf121} we can clearly see that the AST model shows the same performance from the BirdCLEF 2022 challenge. As mentioned in the methods section, the BirdCLEF 2021 challenge differs from this \textit{years challenge due to the availability of training data}. We had 62,000 bird calls from all around the world whereas in this years challenge we only have 15,000 bird calls. This is a \textit{reduction in the number of training data available for each of the bird classes and even then, previous years winners used techniques to combat weak supervision.} However, for the AST model we did not have to use any such technique to yield good results. Another interesting conclusion that can be drawn is that the AST model actually overfit on certain classes of the training data. This can be viewed in the samples F1 score on the validation set. The Convolution baseline seems to have not overfit. This can be viewed in the table \ref{tab:samplesf121}. We could have fixed this issue by setting a higher threshold for early stopping. This can also be honed to the fact that the Macro F1 score and Samples F1 paint two different picture, unlike the previous experiment on BirdCLEF 2022. Therefore, \textit{we see an expected behaviour in the transformer based model as compared to the convolutional baseline.}

From \ref{fig:AST21} we can see the training curves corresponding to the table \ref{tab:macrof121}. The similar trend is observed here as well. The AST model starts to converge after 15 epochs whereas the convolutional baseline \ref{fig:EB21} converges at around 35 epochs. Another interesting conclusion we can draw from this training graphs is that in \ref{fig:EB21} we can see that the training occurred till 40 epochs, however for the AST model the training stopped at 36 epochs. This explains why the model finished training quicker than the convolutional baseline model. We could have let the models train for longer i.e. more epochs but we hit the GPU limitations on Kaggle. Another conclusion we have is that the model took longer to train is due to the size of the dataset. 

The loss curves are similar to the experiments we ran on BirdCLEF22. This can be visualized in \ref{fig:AST21}, \ref{fig:EB21} and \ref{tab:losses21}. We can see that the AST models has the lowest loss and the model loss keeps steadily decreasing after a initial jump. 

\subsection{Discussion: ``Macro" vs ``Samples"}

The Macro F1 score collects a general average over all classes and hence if performance is not consistent over and only does well for some of the classes the scores would be lower. Considering we have 152 classes, and the performance would be expected to be spread out based on their distribution in training data the Macro F1 score is relatively lower almost by an order.

From figure \ref{fig:dist} we know many of the 152 classes have low recordings, impacting the Macro score. 

\subsection{Future Work}
Since our set of experiments are limited (5 models and 2 datasets), some questions remain unanswered. 

First is \textit{Does ensembling, which worked in previous edition of BirdCLEF\cite{CondeEtAl:CLEF-2021}\cite{HenkelEtAl:CLEF-2021} with of convolutional backbone help in transformer backbone models? If it does,  does the \textit{difference} in perfomance between the two hold?}

Second is \textit{What causes the difference in performance between the convolutional backbones and transformer backbone?} specifically \textit{what classes are harder for each of these.} This would help us be more cognizant of our backbone choice and it's impact when finally utilizing them for the purpose of bird sound identification. Third leads from the second. \textit{Can joint ensembling of the Transformer and Convolution baselines, help us improve overall performance by utilizing the predictions of the two different architectures, based on their specific strengths?}

The natural line of research is how can we \textit{improve the usability of these models on edge devices}. Edge devices can be smart phones or IoT devices like Raspberry Pi, that can be setup and installed on forest floors or trees. These devices can be recording and monitoring bird calls live and provide updates to scientists if they are in a look out for a particular species.
% Model parameters
% 4201684
% 311448
% 87484264
% Conclusions - summarize in one paragraph what is the main take-away point from your work. Add a final paragraph discussing any potential ethical implications of your project (e.g., fairness, accountability, transparency, privacy, social impact, etc).
\section{Conclusion}

Our experiments have shown that the Audio Spectrogram Transformer baseline outperforms the convolutional baseline. The AST which had been trained on audio sounds represented as spectograms, showed better performance than conventional visual style pre-trained EfficientNet. The goal of this was for recommendation of backbone for the BirdCLEF tasks. Using AST instead of convolutional backbones, may provide better performance in the task going forward. Even if it is not unequivocal, AST has to be is a \textit{genuine choice} for the task.

This work has a positive environmental and ethical impact intention, since the task central to it aims at aiding conservation. With the dwindling population of birds we are using deep learning in the most responsible way to save these birds, while also preserving and maintaining their records. The future scope and line of research ties in with the ethical transparency, as we can open source this models and build an application that can be used by hundreds of folks around the world to help gather and pool in the data on these birds migratory paths and sustenance. We also note that, deep learning methods have known to be compute and energy intensive and their proportional emission cost cannot be neglected\cite{strubell-etal-2019-energy}, including in our own experiments. Our work helps in answering a design choice, but tools built should undergo wider testing and analysis of trade-offs, including by domain experts.

Our work does not propose a tool that can replace human experts. It acts as knowledge for the designers of bird sound identification tool, and should be interpreted as such.

\bibliographystyle{dinat}
\bibliography{egbib}

\end{document}